\documentclass[aps,prb,reprint,preprintnumbers,amsmath,amssymb,floatfix,longbibliography]{revtex4-1}

\usepackage{graphicx}
\usepackage{dcolumn}
\usepackage{bm}
\usepackage{hyperref}
\usepackage{float}

\begin{document}
\topmargin -4 mm

\title{ Is a COVID19 Quarantine Justified in Chile or USA Right Now?  }

\author{R. I. Gonzalez}
\altaffiliation{Center for the Development of Nanoscience and Nanotechnology}
\affiliation{Centro de Nanotecnolog\'ia Aplicada, Universidad Mayor, Santiago, Chile}
\email{m.kiwi.t@gmail.com and rafael.gonzalez@umayor.cl}

\author{F. Munoz}%
\altaffiliation{Center for the Development of Nanoscience and Nanotechnology}
\affiliation{Departamento de F\'isica, Facultad de Ciencias, Universidad de Chile, Santiago, Chile.}

\author{P. S. Moya}
\affiliation{Departamento de F\'isica, Facultad de Ciencias, Universidad de Chile, Santiago, Chile.}

\author{M. Kiwi}
\altaffiliation{Center for the Development of Nanoscience and Nanotechnology}
\affiliation{Departamento de F\'isica, Facultad de Ciencias, Universidad de Chile, Santiago, Chile.}

\date{\today}

\begin{abstract}
During the current COVID-19 pandemic it is imperative to give early warnings to reduce mortality. However, non-specialist such as authorities and the general population face several problems to understand the real thread of this pandemic, and under/overestimation of its risk are a commonplace in the press and social media. Here we define an index, which we call the COVID-19 Burden Index, that relates the capacities of the healthcare system of a given country to treat severe and critical cases. Its value is 0 if there is no extra strain in the healthcare system, and it reaches 1.0 when the collapse is imminent. As of 23 March 2020, we show that Chile, the USA, UK, among other countries, must reduce the rate of infections right now, otherwise, in less than 7 days they could be in a catastrophic situation such as Italy, Spain and Iran.
\end{abstract}

\maketitle

Is a nationwide quarantine, or other harder measures justified in
Chile? The answer without deeper analysis is yes. But, the real tricky
ingredient is timing.  Scientific evidence has to be a key element in
making this decision, and it has to be based on the available data. A
reckless decision could do enormous harm to the national economy,
 which in Chile is already crippled by the social outburst of
October 2019. However, acting too late is likely to imply a large
number of causalities, or producing an even harder sociopolitical
scenario. The long delay between the enforcement of measures and their
repercussion in the infection rate makes taking the right decision
even more difficult.

Our intention is to show, with just one -easy to grasp- parameter, how
bad the situation facing the national healthcare system is. And, on that basis, infer if a nationwide quarantine should be
enforced as soon as possible. We avoid making forecasts about the
number of infected, limiting ourselves to the available data. Our
assumptions are minimal and play almost no role in our conclusions. We
suggest a nationwide quarantine considering that our social life is
very similar to that of Italy or Spain, who are experiencing a
giant humanitarian crisis due to the collapse of their healthcare
system. But, beyond this or another measure, we want to show that the
contagion of the disease must be radically reduced right now, and this
can be achieved by effective social distancing, only possible by
lockdown.

\section{The COVID-19 burden index}

In an epidemic scenario, rather than the contagion percentage of the
total population, one of the most critical elements is the strength
of the healthcare system. In the COVID-19 pandemic, the bottleneck of
the healthcare system is given by the number of beds
\textit{available} for intensive care (ICU). The number of ICU beds in
Chile is not entirely certain, but it could be estimated to 1000
beds\cite{camasChile} (or $\sim 5.3$ ICU per 100.000 inhabitants). Of
those, we can expect a steady-state occupancy of 75\%; this is the
average of the usual hospital beds occupied in the
OECD~\cite{ocde_icu}.
Due to the exponential growth of the infection, a lower value
(\textit{i.e.}, a 50\% or a 30\% occupancy) does not make much
difference. Therefore, for this study, we are going to consider the
currently available capacity of the Chilean healthcare system for
COVID-19 patients is $\sim 250$ ICU beds. As we will show later, a
higher number of ICU beds can be assumed, for example, similar to the
case of Belgium, but the scenario is still quite worrying.


The ratio of COVID-19 infected to those in need of intensive or
critical care is about 15\%, and a third of them require
mechanical ventilation~\cite{liew2020,arabi2020,xie2020l,hopman2020}. We will use
15\% for our estimate.  To compare the evolution in different
countries we define the \textbf{COVID-19 burden} index as
\begin{equation*}
    \frac{0.15\times N_{\text{confirmed cases}}}{0.25\times N_{\text{beds}}},
\end{equation*}
\noindent where $N_{\text{confirmed cases}}$ the number of cumulative
confirmed cases, and $N_{\text{beds}}$ the number of ICU beds
available in each country prior to 2020~\cite{camasMundo}. A zero
value means the basal occupancy of the healthcare system, {\it i.e.},
the healthcare system in steady-state before the year 2020.  When this
index reaches 1.0 (the Burnout Limit) the healthcare system is at its
local limit. A value larger than 1.0 reflects, approximately, how many
extra ICU and medical professionals are needed relative to the regular
healthcare system situation (75\% occupancy). The important message is
to keep the COVID-19 burden index below unity. For values above one 
doctors will surely find themselves in the difficult humanitarian
dilemma of who to care for, or even worse, who to save.

\begin{figure}[H]
\includegraphics[width=0.95\columnwidth]{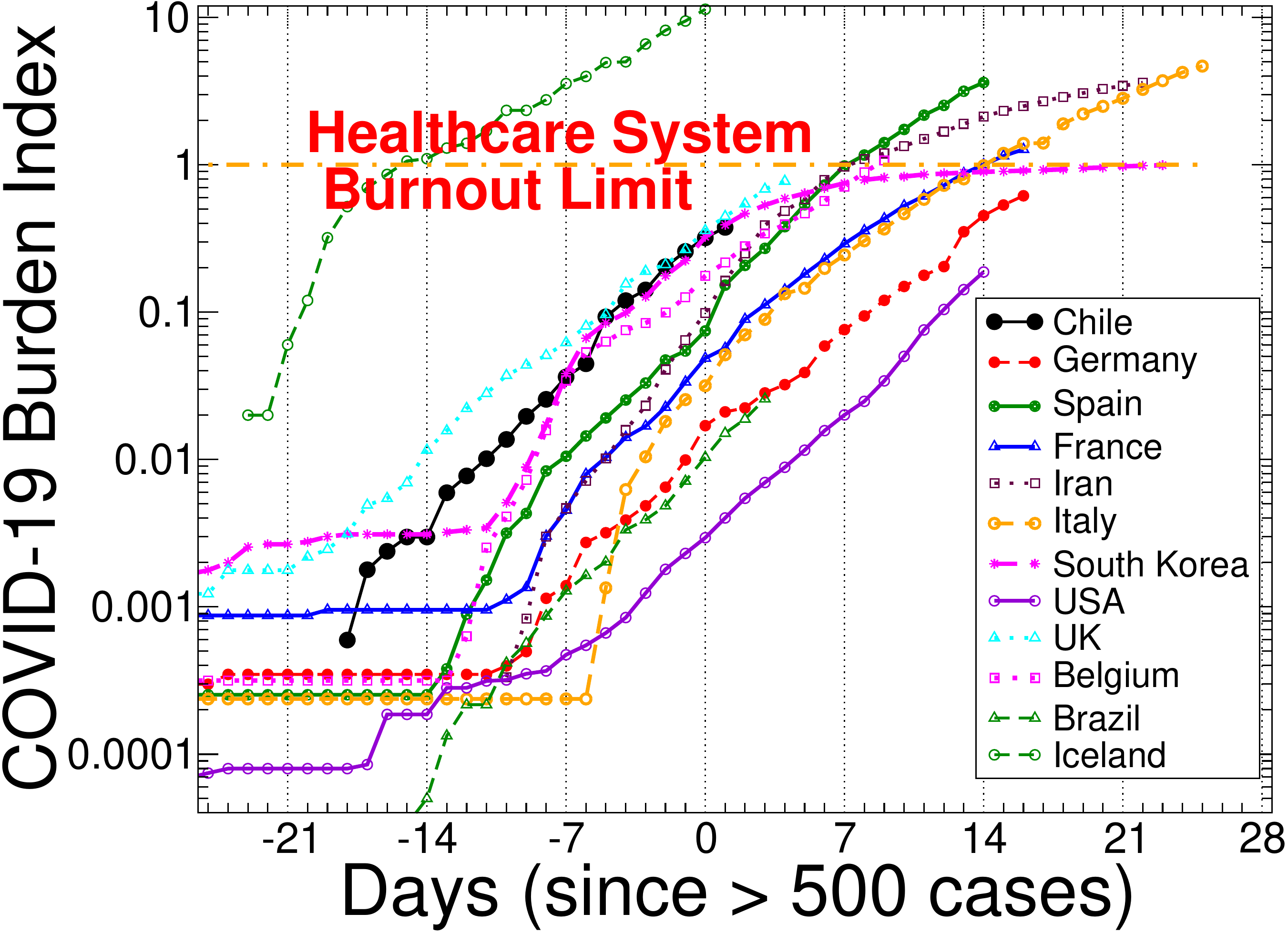} 
\caption{\label{fig:HIndex} COVID-19 burden index for several
  countries as a function of time, after the first 500 positive cases are
  confirmed, a reasonable value for Chile. Notice that we present the
  index on a logarithmic scale.}
\label{fig:burden-index}
\end{figure}

\begin{figure}[h]
\includegraphics[width=\columnwidth]{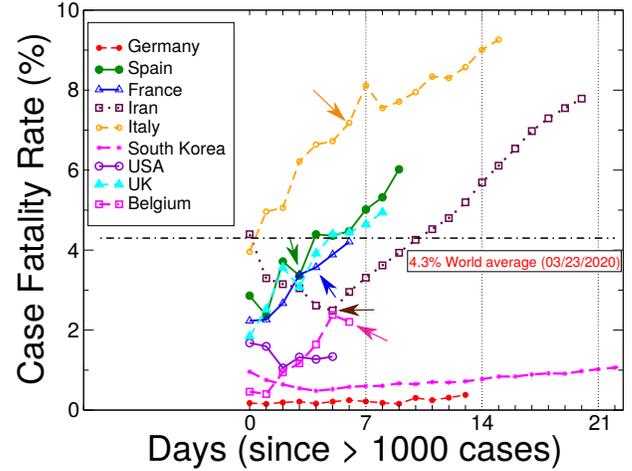} 
\caption{\label{fig:fatality}Fatality rate of COVID-19 by
  country. Only the data for countries with more than 1000 infected
  cases is provided.  The black line is the world average by March 23,
  2020. The small arrow indicates when the Burnout Limit was reached
  (COVID-19 Burden Index of 1). }
\label{fig:cfr}
\end{figure}

Figure~\ref{fig:burden-index} shows the COVID-19 burden index for
several countries using the data available \cite{covidData}. The
results are consistent with what is expected; the worst affected
countries, {\it i.e.}, Italy, Iran, and Spain score 4.6, 3.6, and 3.6
in the last day considered for this report. UK and Germany are about
to reach their limit. Germany has maintained a controlled COVID-19
Burden index for 16 days; that is, it has maintained its health system
without collapsing. But 16 days below the burnout limit may be
insufficient, if they do not manage to reduce the rate, it spreads
urgently. Although the disease is overcome on average in 2 weeks,
critically ill patients take between 3 and 6 weeks~\cite{WOH}, so
Germany seems still far from overcoming the emergency. The UK case is
very worrying because it could collapse very quickly if it does not
reduce the rate of infections as soon as 2 or 3 days, or they could
face the worst-case scenario. South Korea is on its limit, its
healthcare system is fully occupied and strained, but it is copping
with the demand. Finally, USA and Chile are steadily approaching the
limits of their healthcare systems; in about one week, they might not
be able to cope with the demand if they don't reduce the rate of
contagion right now. An under/over estimate of the parameters used
could mean at most one or two days in reaching the burnout
limit. Close to reaching this limit, the fatality of COVID-19
increases drastically, see Fig.~\ref{fig:fatality}.

\begin{figure}[h]
\includegraphics[width=\columnwidth]{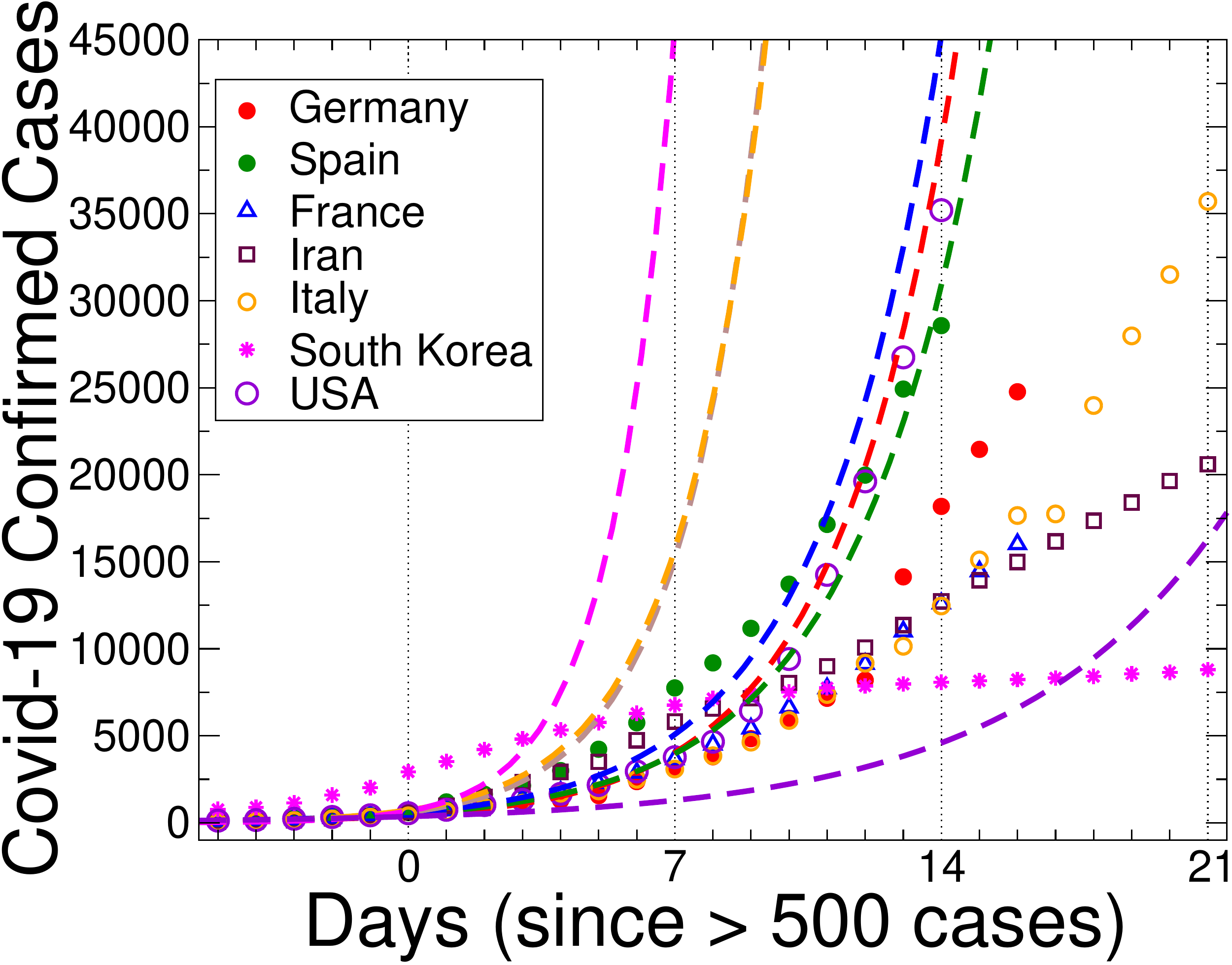}
\caption{\label{fig:pred}Is it possible to predict the contagion trend
  with data of about 500 cases? An exponential trend (dashed lines)
  was adjusted for each country real data using values over 50 and just
  over 500 cases. As a comparison, the actual data for the full period is
  plotted as symbols. Despite the common belief of decision-makers and
  the general population, it is NOT possible to give a truthful
  prediction, for better or for worse scenarios, during the early
  stages of COVID-19 spread based exclusively on the number of
  confirmed cases.}
\end{figure}

Looking to the slope of the COVID-19 burden index in a logarithmic
scale from countries such as Iceland, Belgium, Chile, France, Germany,
and USA, all of them have similar behavior. At \textit{early}
stages, the contagion depends much on details, such as population,
demographics and/or geography. Then, at \textit{intermediate} stages
-when around the 500 detected cases is reached- the slope of the
COVID-19 burden index is nearly the same. This applies to countries so
different as Iceland, Chile, USA and South Korea.  It seems that once
the exponential growth of the epidemic is unleashed, specific
details for each country become irrelevant. Also, this slope does not
correlate with the contagion rate found during the early propagation
stages. Afterward, at \textit{late} stages, the slope decreases. This
is the case of South Korea, Italy and Iran. This third stage, we infer,
is related to contagion control reducing the contagion rate. But,
if this is accomplished too late as in Iran or Italy, the
consequences are disastrous, as is being experienced. Our
recommendation, for any country that is still in time, is to stay under
the burnout point (1.0) as South Korea achieved. As we will show
below, this can make a difference in reducing the number of deaths,
when compared to the Italian crisis by about 10 times, using the
preliminary data from South Korea, or even up to about 20 times as
preliminary data from  Germany. Finally, the marked
difference, between 3 clear stages (or more in the case of South
Korea) makes a prediction of the spread of COVID-19 inaccurate using
standard methods, such as a regression of early contagious data (see
Fig.~\ref{fig:pred}).

The most significant danger of this pandemic is not how contagious it is,
or the mortality rate, which are important. The main danger is the
collapse of the healthcare system. The case fatality rate (CFR), defined
as the percentage of death from confirmed COVID-19 cases, changes day
to day, and cannot be determined with certainty, and our data are no
reference. However, on the basis of the data available on March 23, it
is around 4,3\%. This is in sharp contrast with South Korea, with over
8,000 cases and about 1\% CFR. However, Spain and Italy, with their
collapsed healthcare systems, report around 6 and 9\%,
respectively. Germany has not overcome the emergency and must take
extreme measures. In any case, having today a similar number of
confirmed infections than those of Spain, and after a similar 
time since they both exceeded 500 cases, it has a CFR of less than
0.5\%, which seems encouraging.

Mainly, our intention is to sound an alarm. It is vital that the
authorities of every affected country, on the basis of ICU available
beds and the contagion rate, make their most to keep the COVID-19
Burden Index below 1, and take the severe and difficult decisions that
this objective requires. In this analysis, we have tried to avoid any
bias, sticking to the available data~\cite{covidData}. We believe that
the COVID-19 burden index is a simple and solid tool to warn the
authorities and decision-makers about the dangers of COVID-19. If the
COVID-19 burden index reaches values close to 1.0 the collapse of the
healthcare system is imminent, and harsh measures are in need to avoid
a drastic increase in causalities.

Finally, while our main concern is Chile, the US case is very
worrisome. In fact, with more than 100,000 ICU beds, twice as much as
China, they are not shielding the population from exponential
COVID-19 growth. The US has 17 times more population than Chile, and
100 times more ICU beds, but there is no guarantee of avoiding
collapse (see Fig.~\ref{fig:HIndex}). The only way to avoid collapse
is by reducing the contagion rate as early as possible. By March 23,
the slope of the COVID-19 burden index of USA is similar to South
Korea in Day 2 (see Fig.~\ref{fig:HIndex}). South Korea shortly
afterward decreased the slope of the curve, avoiding the burnout
limit. The same could happen in USA, but it will depend on the
measures taken in the previous weeks.

The authors of this brief communication are used to analyze data and
perform simulations in physics, trying to describe
Nature. Regrettably, in this urgent communication, we rely only on real
data and try to find a simple interpretation of a global humanitarian
crisis to help on its solution. The scientific research using
computational simulations and experiments are essential to stop this
pandemic. Nevertheless, for this communication, the computational
simulations were not necessary at all; the evidence from several
countries is overwhelming. Unless the exponential growth is hindered
in the following days, Chile, USA, and other countries in a similar
situation could experience a similar collapse of their
healthcare systems. Sadly, to stop the number of critical cases during
the next week, the measures should have been taken about one week
before writing this communication.

\begin{acknowledgments}
The authors are supported by ANID Chile through Fondecyt grants
Nos. 1191351, 1191353, and 11180557, the Center for the Development of
Nanoscience and and Nanotechnology CEDENNA AFB180001, and Conicyt
PIA/Anillo ACT192023.
\end{acknowledgments}

%

\begin{thebibliography}{9}%
\makeatletter
\providecommand \@ifxundefined [1]{%
 \@ifx{#1\undefined}
}%
\providecommand \@ifnum [1]{%
 \ifnum #1\expandafter \@firstoftwo
 \else \expandafter \@secondoftwo
 \fi
}%
\providecommand \@ifx [1]{%
 \ifx #1\expandafter \@firstoftwo
 \else \expandafter \@secondoftwo
 \fi
}%
\providecommand \natexlab [1]{#1}%
\providecommand \enquote  [1]{``#1''}%
\providecommand \bibnamefont  [1]{#1}%
\providecommand \bibfnamefont [1]{#1}%
\providecommand \citenamefont [1]{#1}%
\providecommand \href@noop [0]{\@secondoftwo}%
\providecommand \href [0]{\begingroup \@sanitize@url \@href}%
\providecommand \@href[1]{\@@startlink{#1}\@@href}%
\providecommand \@@href[1]{\endgroup#1\@@endlink}%
\providecommand \@sanitize@url [0]{\catcode `\\12\catcode `\$12\catcode
  `\&12\catcode `\#12\catcode `\^12\catcode `\_12\catcode `\%12\relax}%
\providecommand \@@startlink[1]{}%
\providecommand \@@endlink[0]{}%
\providecommand \url  [0]{\begingroup\@sanitize@url \@url }%
\providecommand \@url [1]{\endgroup\@href {#1}{\urlprefix }}%
\providecommand \urlprefix  [0]{URL }%
\providecommand \Eprint [0]{\href }%
\providecommand \doibase [0]{https://doi.org/}%
\providecommand \selectlanguage [0]{\@gobble}%
\providecommand \bibinfo  [0]{\@secondoftwo}%
\providecommand \bibfield  [0]{\@secondoftwo}%
\providecommand \translation [1]{[#1]}%
\providecommand \BibitemOpen [0]{}%
\providecommand \bibitemStop [0]{}%
\providecommand \bibitemNoStop [0]{.\EOS\space}%
\providecommand \EOS [0]{\spacefactor3000\relax}%
\providecommand \BibitemShut  [1]{\csname bibitem#1\endcsname}%
\let\auto@bib@innerbib\@empty
\bibitem [{\citenamefont {de~Salud~de Chile}()}]{camasChile}%
  \BibitemOpen
  \bibfield  {author} {\bibinfo {author} {\bibfnamefont {M.}~\bibnamefont
  {de~Salud~de Chile}},\ }\href@noop {} {\bibinfo {title} {Informe campa\~na de
  invierno 2017}},\ \bibinfo {howpublished}
  {\url{https://www.minsal.cl/wp-content/uploads/2018/03/Informe-Final-Campa\%C3\%B1a-de-Invierno-2017.pdf}}\BibitemShut
  {NoStop}%
\bibitem [{\citenamefont {OECD}()}]{ocde_icu}%
  \BibitemOpen
  \bibfield  {author} {\bibinfo {author} {\bibnamefont {OECD}},\ }\href@noop {}
  {\bibinfo {title} {Hospital beds and discharge rates}}\BibitemShut {NoStop}%
\bibitem [{\citenamefont {Liew}\ \emph {et~al.}(2020)\citenamefont {Liew},
  \citenamefont {Siow}, \citenamefont {MacLaren},\ and\ \citenamefont
  {See}}]{liew2020}%
  \BibitemOpen
  \bibfield  {author} {\bibinfo {author} {\bibfnamefont {M.~F.}\ \bibnamefont
  {Liew}}, \bibinfo {author} {\bibfnamefont {W.~T.}\ \bibnamefont {Siow}},
  \bibinfo {author} {\bibfnamefont {G.}~\bibnamefont {MacLaren}},\ and\
  \bibinfo {author} {\bibfnamefont {K.~C.}\ \bibnamefont {See}},\ }\bibfield
  {title} {\bibinfo {title} {Preparing for covid-19: early experience from an
  intensive care unit in singapore},\ }\href@noop {} {\bibfield  {journal}
  {\bibinfo  {journal} {Critical Care}\ }\textbf {\bibinfo {volume} {24}},\
  \bibinfo {pages} {1} (\bibinfo {year} {2020})}\BibitemShut {NoStop}%
\bibitem [{\citenamefont {Arabi}\ \emph {et~al.}(2020)\citenamefont {Arabi},
  \citenamefont {Murthy},\ and\ \citenamefont {Webb}}]{arabi2020}%
  \BibitemOpen
  \bibfield  {author} {\bibinfo {author} {\bibfnamefont {Y.~M.}\ \bibnamefont
  {Arabi}}, \bibinfo {author} {\bibfnamefont {S.}~\bibnamefont {Murthy}},\ and\
  \bibinfo {author} {\bibfnamefont {S.}~\bibnamefont {Webb}},\ }\bibfield
  {title} {\bibinfo {title} {Covid-19: a novel coronavirus and a novel
  challenge for critical care},\ }\href@noop {} {\bibfield  {journal} {\bibinfo
   {journal} {Intensive care medicine}\ ,\ \bibinfo {pages} {1}} (\bibinfo
  {year} {2020})}\BibitemShut {NoStop}%
\bibitem [{\citenamefont {Xie}\ \emph {et~al.}(2020)\citenamefont {Xie},
  \citenamefont {Tong}, \citenamefont {Guan}, \citenamefont {Du}, \citenamefont
  {Qiu},\ and\ \citenamefont {Slutsky}}]{xie2020l}%
  \BibitemOpen
  \bibfield  {author} {\bibinfo {author} {\bibfnamefont {J.}~\bibnamefont
  {Xie}}, \bibinfo {author} {\bibfnamefont {Z.}~\bibnamefont {Tong}}, \bibinfo
  {author} {\bibfnamefont {X.}~\bibnamefont {Guan}}, \bibinfo {author}
  {\bibfnamefont {B.}~\bibnamefont {Du}}, \bibinfo {author} {\bibfnamefont
  {H.}~\bibnamefont {Qiu}},\ and\ \bibinfo {author} {\bibfnamefont {A.~S.}\
  \bibnamefont {Slutsky}},\ }\bibfield  {title} {\bibinfo {title} {Critical
  care crisis and some recommendations during the covid-19 epidemic in china},\
  }\href@noop {} {\bibfield  {journal} {\bibinfo  {journal} {Intensive Care
  Medicine}\ ,\ \bibinfo {pages} {1}} (\bibinfo {year} {2020})}\BibitemShut
  {NoStop}%
\bibitem [{\citenamefont {Hopman}\ \emph {et~al.}()\citenamefont {Hopman},
  \citenamefont {Allegranzi},\ and\ \citenamefont {Mehtar}}]{hopman2020}%
  \BibitemOpen
  \bibfield  {author} {\bibinfo {author} {\bibfnamefont {J.}~\bibnamefont
  {Hopman}}, \bibinfo {author} {\bibfnamefont {B.}~\bibnamefont {Allegranzi}},\
  and\ \bibinfo {author} {\bibfnamefont {S.}~\bibnamefont {Mehtar}},\
  }\bibfield  {title} {\bibinfo {title} {Managing covid-19 in low-and
  middle-income countries},\ }\href@noop {} {\bibinfo  {journal} {JAMA}\
  }\BibitemShut {NoStop}%
\bibitem [{cam()}]{camasMundo}%
  \BibitemOpen
\bibfield  {journal} {  }\href@noop {} {}\bibinfo {howpublished}
  {\url{https://en.wikipedia.org/wiki/List_of_countries_by_hospital_beds}}\BibitemShut
  {NoStop}%
\bibitem [{\citenamefont {Max~Roser}\ and\ \citenamefont
  {Ortiz-Ospina}(2020)}]{covidData}%
  \BibitemOpen
  \bibfield  {author} {\bibinfo {author} {\bibfnamefont {H.~R.}\ \bibnamefont
  {Max~Roser}}\ and\ \bibinfo {author} {\bibfnamefont {E.}~\bibnamefont
  {Ortiz-Ospina}},\ }\bibfield  {title} {\bibinfo {title} {Coronavirus disease
  (covid-19) – statistics and research},\ }\href@noop {} {\bibfield
  {journal} {\bibinfo  {journal} {Our World in Data}\ } (\bibinfo {year}
  {2020})},\ \bibinfo {note}
  {https://ourworldindata.org/coronavirus}\BibitemShut {NoStop}%
\bibitem [{\citenamefont {Organization}(2020)}]{WOH}%
  \BibitemOpen
  \bibfield  {author} {\bibinfo {author} {\bibfnamefont {W.~H.}\ \bibnamefont
  {Organization}},\ }\href@noop {} {\bibinfo {title} {Report of the who-china
  joint mission on coronavirus disease 2019 (covid-19)}},\ \bibinfo
  {howpublished}
  {\url{https://www.who.int/docs/default-source/coronaviruse/who-china-joint-mission-on-covid-19-final-report.pdf}}
  (\bibinfo {year} {2020})\BibitemShut {NoStop}%
\end{thebibliography}

\end{document}